\documentclass[twocolumn,showpacs,aps,superscriptaddress]{revtex4}
\usepackage{latexsym}
\usepackage[dvips]{graphicx}

\begin{document}

\title{Finite-Size effects in "Single Chain Magnets": an experimental and theoretical study}
\author{L. Bogani}
\affiliation{Dipartimento di Chimica and INSTM, Universit\`a di
Firenze,  I-50019 Sesto Fiorentino, Italy}
\author{A. Caneschi}
\affiliation{Dipartimento di Chimica and INSTM, Universit\`a di
Firenze,  I-50019 Sesto Fiorentino, Italy}
\author{M. Fedi}
\affiliation{Dipartimento di Fisica and u.d.r. INFN, Universit\`a di
Firenze,  I-50019 Sesto Fiorentino, Italy}
\author{D. Gatteschi}
\affiliation{Dipartimento di Chimica and INSTM, Universit\`a di
Firenze,  I-50019 Sesto Fiorentino, Italy}
\author{M. Massi}
\affiliation{Dipartimento di Fisica and u.d.r. INFN, Universit\`a di
Firenze,  I-50019 Sesto Fiorentino, Italy}
\author{M.A. Novak }
\affiliation{Instituto de Fisica Universidade Federal do Rio de Janeiro,
CP68528, RJ 21945-970, Brazil}
\author{M.G. Pini}
\affiliation{Istituto di Fisica Applicata "Nello Carrara", CNR,
 I-50019 Sesto Fiorentino, Italy}
\author{A. Rettori}
\affiliation{Dipartimento di Fisica and u.d.r. INFM, Universit\`a di
Firenze,  I-50019 Sesto Fiorentino, Italy}
\author{R. Sessoli}\email[corresponding
author:]{roberta.sessoli@unifi.it}
\affiliation{Dipartimento di Chimica and INSTM, Universit\`a di
Firenze,  I-50019 Sesto Fiorentino, Italy}
\author{A. Vindigni}
\affiliation{Dipartimento di Fisica and u.d.r. INFM, Universit\`a di
Firenze,  I-50019 Sesto Fiorentino, Italy}

\date{30 March 2004}

\begin{abstract}
The problem of finite size effects in s=1/2 Ising systems showing slow dynamics of the
magnetization is investigated introducing diamagnetic impurities in a Co$^{2+}$-radical chain. The static magnetic
properties have been measured and analyzed considering the peculiarities induced by the
ferrimagnetic character of the compound. The dynamic susceptibility shows that an Arrhenius law is observed with the
same energy barrier for the pure and the doped compounds while the prefactor decreases, as theoretically predicted.
Multiple spins reversal has also been investigated.
 \end{abstract}


\maketitle

One of the open problems in low dimensional magnetism is the influence of finite-size effects\cite{Xu,Eggert} and
investigation of their role on dynamic properties,
 which are of great relevance for nanostructures, is particularly interesting.
For the Ising spin chain Glauber proposed a stochastic dynamics\cite{Glauber}
widely applied to a variety of phenomena.
The experimental investigation of this dynamics in magnetic 1D compounds was prevented by
 the strict requirements the system must fulfill. Recently molecular 1D materials were found to show magnetic
hysteresis in absence of long range
magnetic order\cite{catan,verd,clerac}. They have thus been named "Single Chain Magnets"
(SCMs) by analogy to slow relaxing 0D systems widely known as "Single Molecule Magnets" (SMMs)\cite{RevAng}.
The first of these compounds
was Co(hfac)$_{2}$NitPhOMe (CoPhOMe in the following)\cite{catan}, made of alternating and interacting
Co$^{2+}$ ions (anisotropic effective $s$=1/2 spins)
and  nitronyl-nitroxide radicals (isotropic $s$=1/2 spins).
The relaxation time follows an Arrhenius law as
predicted by Glauber\cite{Glauber} for a 1D Ising system, with
$\tau =\tau_{0}e^{\Delta/k_{B}T}$ and $\Delta/k_{B}=$152K,
for ten decades of time in the temperature range $T=$4-15K \cite{EPL}.

The observation of Glauber dynamics in the paramagnetic phase of a real system represents
a paradox. In fact a strong Ising type intra-chain interaction is required because $\Delta$
is proportional to the coupling constant $J$.
On the other hand the correlation length  of the Ising model diverges exponentially
 at low temperature, $\xi\sim exp(\frac{2J}{k_{B}T})$,
and favors the 3D ordering by enhancing the weak inter-chain interactions \cite{Hone}.
A possible explanation could be the presence of naturally occurring defects, which
can geometrically limit the correlation length and dramatically reduce the 3D ordering temperature
 \cite{Hone,Dupas}.
The presence of defects is known to affect also the dynamics of the magnetization, but direct investigations
of their influence on the slow relaxing magnetization are still lacking. The
persistence of slow dynamics even for small segments of chains
is an important result, making SCMs much more appealing than SMMs in many respects.
Finite-size effects are commonly investigated through the insertion of breaks by doping with diamagnetic ions, but
important aspects, like the reproducibility or homogeneity of the concentration of the dopant, are often overlooked.
We present here a thorough experimental and theoretical
investigation of finite-size effects in the slow relaxing Ising chain CoPhOMe, thus evidencing the key role played
 by the defective sites in the nucleation of the excitation, similarly to what is observed in quasi-1D oxides \cite{Xu}.
 A short-cut of the Glauber barrier through coherent reversal of multiple spins
 is observable in the ac susceptibility thanks to the introduction of defects.

 Variable concentrations
of diamagnetic impurities have been inserted in the chains and the samples
 have been investigated with unprecedented accuracy.
Doped crystals were prepared \cite{Dalton}
adding to the solutions different ratios of  Zn(hfac)$_{2}\cdot2$H$_{2}$O vs Co(hfac)$_{2}\cdot$2H$_{2}$O in the
range 0-0.30. As a strongly inhomogeneous distribution of the dopant inside single
crystals could be of major relevance for the magnetic properties, we
used the Particle Induced X-ray Emission technique (PIXE) with an external
micro-beam \cite{PIXE} to analyze the Zn doping profile.
 The low minimum detection limits of PIXE allowed us to find that no metals other
than Co and Zn were present in concentrations $>$12 ppm
and thus no appreciable contribution could rise from paramagnetic impurities.
 The mean Zn/Co ratios of the investigated samples, $c$,
were 0.3\%, 1.9\%, and 4.7\%, giving an
 average length $\overline{L}$ between two Zn$^{2+}$ ions variable from ca. 300 to ca. 20 spins.
 The Zn concentration was always found to be lower than the ratio of the starting solution
and reproducible in crystals from the same batch (within an error of 5\% on $c$)
 but not reproducible in different syntheses.
 Using the micro-beam we could then monitor the spatial
distribution of the dopant inside individual crystals, a procedure never performed
before in this kind of studies.
 The results are shown in Fig.1 for longitudinal and transversal scans. The former shows
uniformly distributed impurities while the latter shows that
 the Zn concentration slightly increases going from the center to the edges. This observation
  is consistent with a progressive enriching of the solution in Zn content.
    \begin{figure}[t]
  \centering
  \includegraphics*[bb=0 80 760 580,scale=0.26]{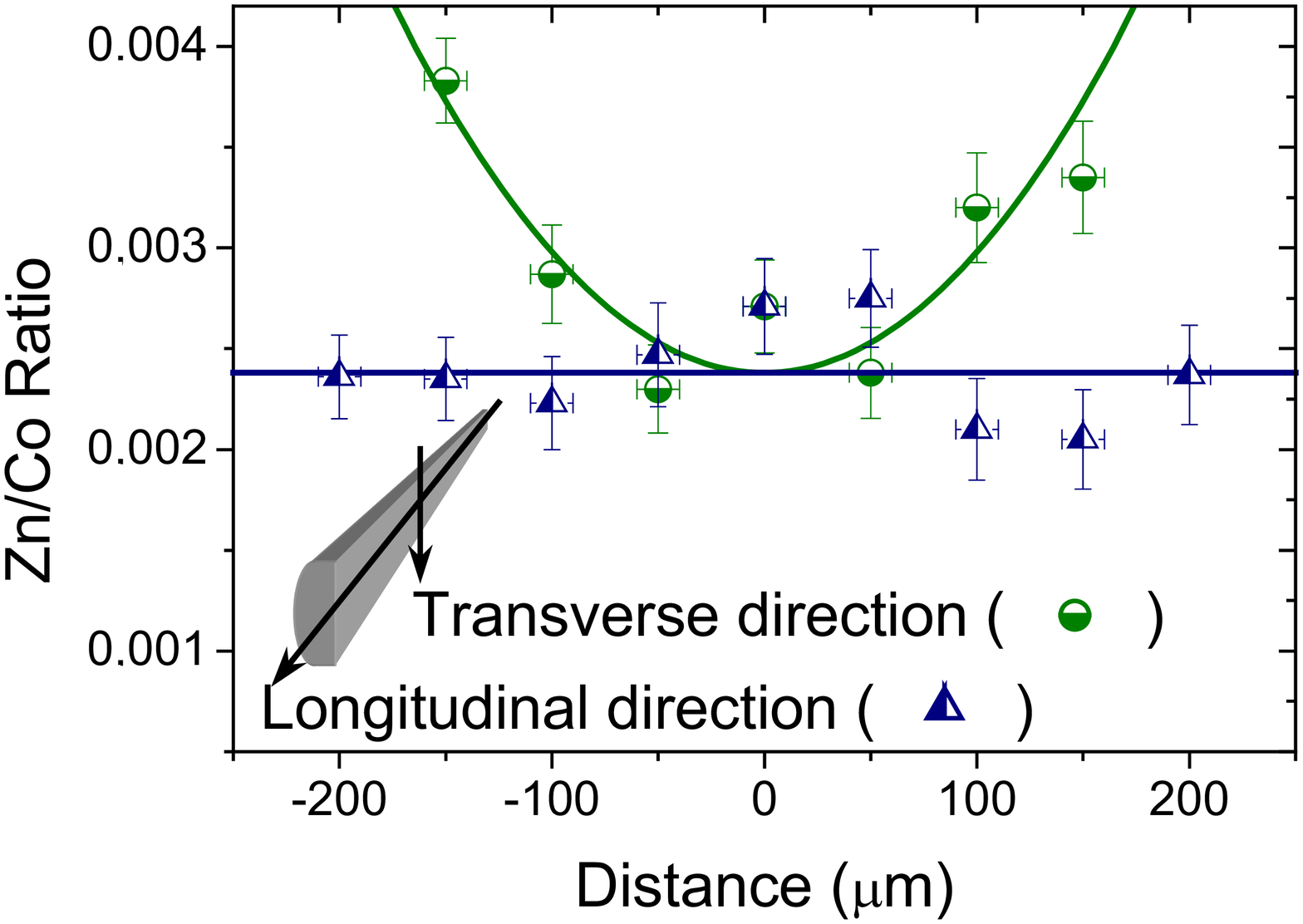}
  \caption{(Color online) Zn/Co ratio in a single crystal of CoPhOMe.
  The distribution of the dopant is reported for both longitudinal and transversal directions,
   scanned as shown in the drawing. Lines are guides to the eye.}
  \label{fig1}
\end{figure}

The magnetic susceptibility of a chain in a moderate field is strongly sensitive to geometrical limitations
of the correlation length. In Fig.2 (top) we report the temperature
dependence of the real component $\chi'$ of the $ac$ magnetic susceptibility
measured at 2.7kHz in a static magnetic field $H=2$kOe along the easy axis for an undoped crystal and
for two doped samples. In contrast with magnetic data and theoretical calculations
 reported so far for chains, which show the presence of only one peak \cite{wortis,Matsubara,RetPini},
  it is evident the presence of a large anomalous
 structure with two peaks for $T\sim33$K and for $T\sim14$K (below the blocking temperature $T_{b}\simeq12$K we observe
 frequency dependent dynamical effects). The higher temperature peak roughly corresponds
 to that expected for an infinite 1D Ising system in presence of a field
 (see Fig.2 bottom) and shifts to higher temperature on increasing the field, as expected.
 This peak is more pronounced in the pure sample and decreases
 and shifts to lower temperatures with higher dopings, while the
 lower temperature peak increases. At high doping (0.047) only this latter peak survives.\begin{figure}[t]
  \centering
  \includegraphics*[bb=50 80 550 730, scale=0.37]{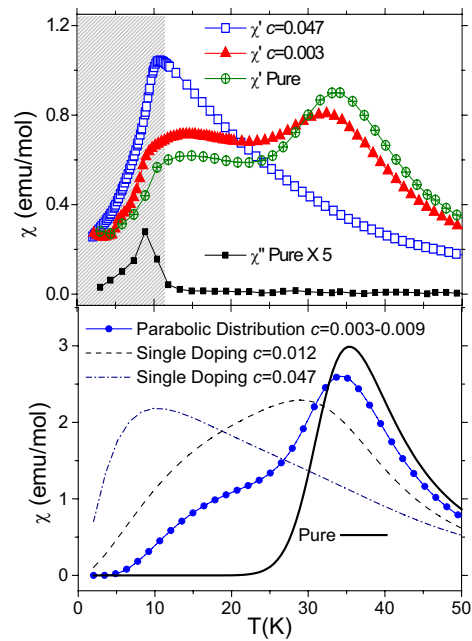}
  \caption{(Color online)(top) Temperature dependence of the longitudinal
   magnetic $ac$ susceptibility (2.7kHz) measured for different dopings in a static field of
   2kOe with an $ac$ field of 10 Oe. Data for other frequencies in the 10 Hz-10kHz range
    are not reported because $\chi^{\prime}$
   is frequency dependent only in the shaded area.
   (bottom) Calculated susceptibility (parameters in the text) for
   different concentrations of impurities. A parabolic distribution is used, at low concentrations, to account
   for the observed transversal trend.}
    \label{fig2}
\end{figure}
 This unprecedented behavior can be explained
 solving the randomly diluted ferrimagnetic Ising chain in
 field. The spin Hamiltonian we have used is:
\begin{equation}
{\cal H}=-\sum_{i=1}^{L/2}[J(\sigma_{2i-1}\sigma_{2i}+\sigma_{2i}\sigma_{2i+1})+\frac{\mu_{B}H}{2}(g_{Co}\sigma_{2i}+g_{R}\sigma_{2i-1})]
\end{equation}
where $g_{Co}$ and $g_{R}$ are the
Land\'{e} factors of Co and radical spins, $\mu_{B}$ is the Bohr magneton,
$H$ is the external magnetic field and $\sigma_{i}=\pm$1, except for $\sigma_{L+1}=0$.
Within a transfer matrix formalism, which is reported
 in literature \cite{wortis,Matsubara,RetPini}, the free energy of a segment
of length $L$ can be obtained. The thermodynamic observables are calculated summing all the $L$'s
weighted by the probability of occurrence of that
length: $P_{L}=c^{2}(1-c)^{L}$ \cite{wortis}.

In Fig.2 (bottom) we report the calculated magnetic susceptibility
 $\chi(T)$ for $J/{k_{B}}$=-90K, $H$=2kOe,
 $g_{Co}$=7 and $g_{R}$=2 and for different doping values.
For the infinite chain we obtain only one peak at
$T\simeq34$K, which shows the observed trend with field, while
 the introduction of non-magnetic impurities gives rise to the
observed peak shift and to a shoulder at about 15K. This feature, which eventually
develops into a single peak at high doping, is mainly due to the finite-size contribution to the free energy\cite{wortis},
 as it will be shown in a forthcoming work.
 Although complete quantitative agreement is not possible due to the helicoidal
 structure of the CoPhOMe chain \cite{EPL,Dalton} the overall behavior and the anomalous feature
are well reproduced by the simple model expressed by (1).
 This is especially true if one introduces, at low concentrations, a parabolic distribution of the dopant,
 as in Fig.1, while at higher concentrations the behavior can be reproduced using either
  a distribution or homogeneous doping.

The presence of the low temperature peak even in the pure compound suggests that
other type of defects are indeed present beyond the introduced Zn impurities. Both
crystalline defects and chemical modifications (due to a certain lability of
the organic radicals) are present in the undoped compound and are expected
 to be more concentrated close to the crystal surface, as for the dopant.
These data suggest a concentration of defects in the pure compound of the order of a few per thousand, a
value that agrees well with the calculated curves of Fig.2 and the following
discussion on the dynamics.

\begin{figure}[t]
   \centering
  \includegraphics*[width=57mm, angle=-90]{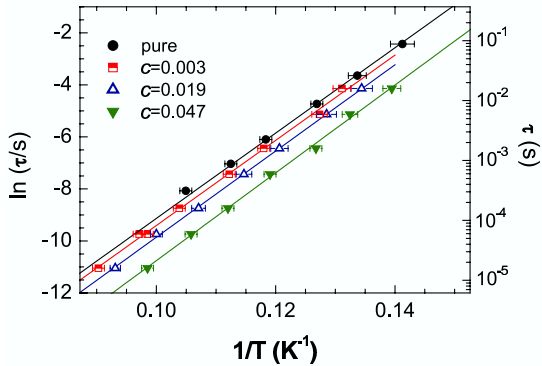}
  \caption{(Color online) Relaxation time of the magnetization extracted from the longitudinal
  $ac$ susceptibility measured in zero static field for different concentrations of Zn.}
  \label{fig3}
\end{figure}

\begin{figure}[t]
  \centering
  \includegraphics*[width=58mm, angle=-90]{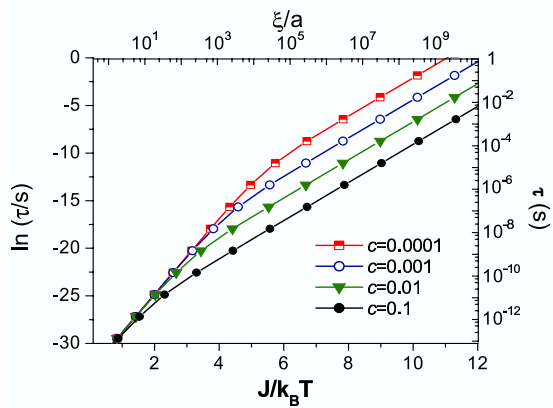}
  \caption{(Color online) Calculated temperature dependence of the relaxation time
  extracted from  $\chi(\omega,T)$ with $\alpha_{0}=3.6\cdot10^{13}$ s$^{-1}$ and $q=0.73$.
   To compare with data in Fig. 3 assume $J/{k_{B}}$=80 K.}
  \label{fig4}
\end{figure}
The dynamics of the magnetization of the pure and doped samples has been
investigated measuring the temperature dependence of the real ($\chi'$) and imaginary ($\chi^{\prime\prime}$)
components of the $ac$ susceptibility for seven different frequencies in the
range 10Hz-10kHz. $\chi^{\prime\prime}$ displays a maximum whose position
depends on the frequency $\omega$ of the applied $ac$ field, and the relaxation time
 can thus be extracted \cite{EPL}: $\tau^{-1}(T_{max})=\omega$.
For the highest concentration the curves become more distorted, with a small shoulder
at low temperature. The results are plotted in Fig.3 in the $\ln(\tau)$ vs 1/$T$ scale.
 A linear behavior is observed for all compounds and the slope (\textit{i.e.}
the barrier $\Delta$ of the Arrhenius law) remains substantially
unchanged ($160\pm8$K). The fitted lines shift down on increasing the doping with the pre-exponential factor
$\tau_{0}$ going from $3.5\cdot10^{-11}$s for the pure compound to $1.0\cdot10^{-12}$s when $c$=0.047.

In order to reproduce theoretically the data of Fig. 3 we assumed the simplified model of a
 dilute ferromagnetic chain (in a ferrimagnet
only an equivalent ferromagnetic branch with $J=|J|$ and $g=g_{Co}-g_{R}$ is
effective for the slow relaxation\cite{EPL}).
For an open chain of length $L$ the time evolution of the  expectation
 values of the spins can be described by the matrix equation:
\begin{equation}
\frac{\partial}{\partial t}\mathbf{S}=-\mathbf{M}\mathbf{S}
\end{equation}
where $\mathbf{M}$ is a tridiagonal matrix.
The time scale of the relaxation process $\tau_{L}(T)$
is related
to the smaller eigenvalue of $\mathbf{M}$.
In the range of $\xi$ and $L \gg 1$ asymptotic expansions for this eigenvalue
are  given in literature, leading to
$\tau_{L}(T)\sim \xi^2$ if $\xi\ll L$, and
$\tau_{L}(T)\sim L \xi$ if $\xi\gg L$\cite{Silva,Lusc}.
However, as the experimental  $\tau$ is extracted from the \emph{ac} susceptibility
of an ensemble of segments with different length, we need to evaluate
$\tau_{L}$ for any $L$ and $T$.
The eigenvalue  spectrum of  $\mathbf{M}$ can be expressed as
\begin{equation}\label{lambda}
\lambda_L(\theta)=\emph{q}\alpha_0\big[1-\gamma cos(\theta)\big]
\end{equation}
where $\emph{q}$ is the probability of reversal of the isolated spin
per unit time  $\alpha_{0}^{-1}$, $\gamma=\tanh(\frac{2J}{k_B T})$, and by $\theta$ we denote the $L$
possible roots of the trigonometric equation 2.6 of ref. \cite{Lusc}.
The relaxation time of the segment of length $L$ is related to the smallest nonzero
root $\theta_{0}$ via $\tau_{L}(T)=\lambda_L^{-1}(\theta_0)$, while the root $\theta=0$
 is rejected since it provides the relaxation time of the infinite chain.
We introduced $\tau_{L}(T)$ in the average susceptibility:
\begin{equation}\label{chi}
\chi(\omega,T)=\frac{\sum_{L=1}^{\infty}P_{L}\chi_{L}(T)(1-\imath\omega\tau_{L}(T))^{-1}}
{\sum_{L=1}^{\infty}P_{L}}
\end{equation}
where $\chi_{L}$ is the susceptibility of a ferromagnetic Ising segment of length $L$ \cite{Matsubara}.
We  extracted the relaxation time of the randomly doped sample
from the maximum of the imaginary component of the computed
$ac$ susceptibility.
In Fig.4 we show the Arrhenius plots thus obtained for different concentrations
 of dopant. A crossover  between the two regions with
$\overline{L}\gg\xi$ and $\overline{L}\ll\xi$ is observed and the slope is halved
in the latter case, where the concentration of impurities \emph{c}
only affects the offset of the curves. This is consistent with a relaxation process
where a domain wall is nucleated at an end point of a segment $L$, with a halved energy cost $2J$,
and has the probability $\frac{1}{L}$
to reach the other end point\cite{Toboshnik}, leading to the linear dependence of the
relaxation time on the size of the system.

The experimental data of Fig.3 scale with  $c$
as the calculated curves when $\overline{L}\ll\xi$, thus evidencing that finite
size effects are also present in the nominally pure compound. Although a direct analysis of its defects
is not possible, by comparing the values of $\tau_{0}$
we estimate an average length of the chains in
 the undoped sample of the order of a thousand spins
(as deduced from the analysis of the susceptibility in a moderate field). The geometrical reduction of $\xi$
can be responsible for the suppression of long
range magnetic order in CoPhOMe \cite{Hone,Dupas}.
The measured activation energy $\Delta$ thus corresponds to
$2J$ instead of $4J$, leading to $J/{k_{B}}=80$ K, in good agreement with
the simulation of the susceptibility data of Fig.2.

Up to now we have neglected a relaxation mechanism which involves the simultaneous
reversal of all the $L$ spins of a segment, whose probability scales as $\emph{q}^{L}$ and therefore is
relevant for short segments. This term can be added to
$\lambda_L(\theta_0)$ to obtain the probability to relax in the unit time $\alpha_{0}^{-1}$.
  \begin{figure}[t]
  \centering
  \includegraphics*[bb=10 90 600 700,width=75mm]{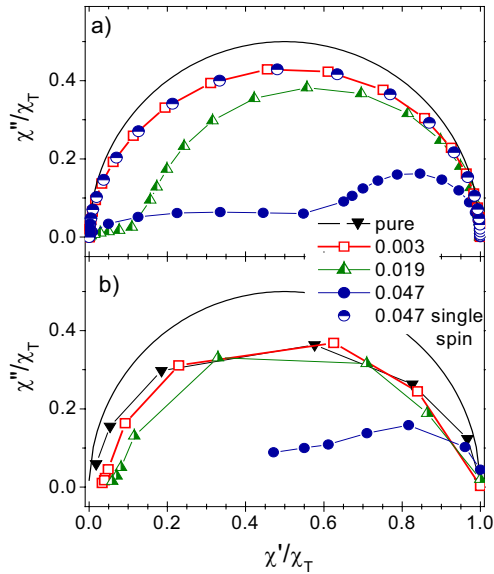}
  \caption{(Color online)(a) Calculated Cole-Cole plot for different concentrations
  of randomly distributed diamagnetic impurities at $T=8$K assuming $J/k_{B}= 80$K,
$q=0.73$, $\alpha_{0}=3.6\cdot10^{13}$ s$^{-1}$. The half-filled circles stay for
 \emph{c}=0.047 without multiple spin reversal.
  (b) Experimental data at $T=8$K for the four investigated samples.}
  \label{fig5}
\end{figure}
Clear evidences of such a relaxation mechanism appear in the Cole-Cole plot (Fig.5a), where
$\chi^{\prime\prime}$ vs. $\chi^{\prime}$ is reported. In the absence of a distribution of relaxation times the
curve is expected to be a semicircle with the center on the $\emph{x}$ axis. A distribution in $\tau$, induced by the random doping,
has  the only effect to slightly push the center down, even for the highest doping. On the contrary, the coherent reversal of all the spins gives rise to a more complex behavior.
The calculated curves of Fig.5a are in fact well below the semicircle and show a flat region for higher frequencies where $\chi^{\prime}\rightarrow0$.
This effect is particularly evident for high doping. The experimental results are shown in Fig.5b. Although the range of available
frequencies is limited, the agreement between observed and calculated
curves suggests that coherent reversal of all the spins is experimentally observed for short segments.
An accurate investigation on the temperature dependence of $\emph{q}$ could discriminate
between a thermally activated and a tunnelling process.

In conclusion we have shown that finite size effects are important in magnetic bistable 1D nanostructures halving
the energy barrier predicted by the Glauber model. We observe a linear dependence of the relaxation time on
the size of the system, which is completely different from the exponential dependence
on the square of the spin value observed in SMMs, or the exponential dependence
on the volume in single domain particles. This opens the possibility to employ segments of chains, even without a precise
control on the length, as nanometric magnetic memory units characterized by a higher
$T_{b}$ and a larger magnetic moment than those obtained in SMMs.
Moreover a non thermally activated relaxation could be observed for short segments.
To better investigate this phenomenon higher doping is required but it cannot be
 obtained with the present chemical approach. Synthesis in confined media, as well as the organization of segments on surfaces, as already done for metallic Co nanowires on Pt surfaces \cite{CoNature},
are under investigation.

Financial support from Italian MIUR (Fondi FISR and FIRB), German DFG, and Brazilian CAPES,
 CNPq and Inst. de Nanociencias is gratefully acknowledged.

\end{document}